\documentclass[fleqn,10pt]{wlscirep}
\title{Fiber-based photon-pair source capable of hybrid entanglement in frequency and transverse mode, controllably scalable to higher dimensions}

\author[1]{D. Cruz-Delgado}
\author[2]{R. Ramirez-Alarcon}
\author[1]{E. Ortiz-Ricardo}
\author[1]{J. Monroy-Ruz}
\author[3]{F. Dominguez-Serna}
\author[1]{H. Cruz-Ramirez}
\author[3,*]{K. Garay-Palmett}
\author[1]{A. B. U'Ren}
\affil[1]{Instituto de Ciencias Nucleares, Universidad Nacional Aut\'onoma de M\'exico, apdo. postal 70-543, 04510 D.F., M\'exico}
\affil[2]{Centro de Investigaciones en \'Optica A.C., Le\'on, Guanajuato 37150, M\'exico}
\affil[3]{Departamento de \'Optica, Centro de Investigaci\'on Cient\'ifica y de Educaci\'on Superior de Ensenada, Apartado Postal 360 Ensenada, BC 22860, M\'exico}
\affil[*]{kgaray@cicese.mx}

\begin{abstract}
We have designed and implemented a photon-pair source,  based on the spontaneous four wave mixing (SFWM) process in a few-mode fiber,  in a geometry which permits multiple, simultaneous SFWM processes, each associated with a distinct combination of transverse modes for the four participating waves.   In our source: i)  each process is group-velocity-matched so that it is, by design, nearly-factorable, and ii)  the spectral separation between neighboring  processes is greater than the marginal spectral width of each process.  Consequently, there is a direct correspondence between the joint amplitude of each process and each of the Schmidt mode pairs of the overall two-photon state.   Our approach permits hybrid entanglement in discrete frequency and in transverse mode, whereby control of the number of supported fiber transverse modes allows scalability to higher dimensions while spectral filtering may be used for straightforward Schmidt mode discrimination.
\end{abstract}
\begin{document}

\flushbottom
\maketitle

\thispagestyle{empty}

\section*{Introduction}

The implementation of photon-pair sources based on the $\chi^{(3)}$ nonlinearity of optical fibers holds much promise for the further advancement of quantum processing technologies~\cite{li05}. Such sources rely on the spontaneous four wave mixing (SFWM) process, in which two pump photons are annihilated leading to the emission of a signal and idler photon pair.  SFWM sources lead to certain distinct advantages when compared to their spontaneous parametric downconversion (SPDC) counterparts, based on  $\chi^{(2)}$ materials.   First, while $\chi^{(2)}$  nonlinear crystals are typically limited to a few cm in length, it is experimentally feasible to implement SFWM sources based on long fibers over which the two pump waves remain temporally overlapped, resulting in longer interaction lengths and consequently higher emission rates.   Second, as discussed in Ref.~\cite{garay07}, SFWM permits a remarkable scope for photon-pair state engineering, for example with a given fiber permitting behaviors ranging from factorable to highly entangled, in contrast to SPDC sources which can be engineered under more restrictive conditions\cite{grice01}.   In this work we exploit a source implemented with a few-mode fiber which can exhibit hybrid entanglement in frequency and transverse mode, and demonstrate a controllable path towards higher-dimensional entanglement. 

It is well known that photon-pair entanglement in continuous degrees of freedom, i.e. frequency-time and transverse position-momentum, can be characterized through a Schmidt decomposition, in which the joint state is written as a discrete sum over factorable contributions of signal and idler Schmidt mode pairs~\cite{eberly00}. Ideally, one would be able to select a given spectral or spatial Schmidt mode for detection, as is routinely done in the case of polarization, e.g. with a polarizer in front of a detector. Such Schmidt-mode discrimination would permit the implementation of quantum information processing protocols with scalability to higher dimensions (as compared to the case of polarization, which is limited to a dimension of $2$). Unfortunately, photon-pair Schmidt modes usually overlap one another making this an experimentally challenging task~\cite{avella14, eckstein11,pinel12}.

We have shown previously that a few-mode, birefringent fiber can permit multiple SFWM processes, each involving a different phasematched combination of transverse modes for pumps 1 and 2, signal, and idler~\cite{cruz14}. A similar approach has been exploited in sources based on the spontaneous parametric downconversion process in $\chi^{(2)}$ nonlinear waveguides~\cite{mosley09,christ09,kruse13,karpinski09}. A distinct advantage of SFWM in few-mode fibers over spontaneous parametric downconversion in few-mode waveguides is that the permitted modes can be very reliably determined in the former, in contrast with the latter; this facilitates accurate modelling of the resulting photon-pair states.   In this paper we extend our earlier work, exploiting the observation that if: i) each joint amplitude, taken individually, is made factorable, and ii) the spectral separation between neighboring SFWM processes is greater than the marginal spectral width of each process, then each joint amplitude in fact constitutes a Schmidt mode pair for the overall two-photon state.   Importantly, the Schmidt modes of the state, which in general exhibits hybrid entanglement in frequency and transverse mode, can be discriminated by straightforward spectral filtering and each Schmidt mode pair can be independently addressed. In this paper we report on the design, implementation, and characterization of a SFWM source which supports three separate processes, each with a nearly-factorable joint spectrum.

The presence of hybrid entanglement implies that two or more degrees of freedom (DOF) are entangled, so that the quantum state cannot be factored into individual states for each DOF.   
Note that this differs with the case of hyperentanglement, for which the state may be written as a tensor product of separate entangled states, one state for each DOF \cite{kwiat97}.   In the context of photon pairs, hybrid entanglement has been demonstrated for polarization and frequency \cite{ramelow09, shu15}, for  polarization and orbital angular momentum \cite{nagali10}, and for  polarization and linear momentum (path) \cite{neves09,ma09}.    From a fundamental point of view, such hybrid systems can serve to prove the independence of entanglement from the specific physical realization of the Hilbert space.   From a practical point of view, hybrid entanglement may be relevant for transmitting entangled pairs over noisy   channels exploiting the ability to choose the DOF with the best robustness for a specific channel\cite{xiao08}.

In this paper we discuss a scheme which permits the emission of photon pairs with hybrid entanglement between \emph{frequency} and \emph{transverse mode}, implemented through the SFWM process in a few-mode fiber.  Note that this type of entanglement may in principle be faithfully transmitted over long distances utilizing  few-mode fibers specifically engineered for the suppression of coupling between the supported modes \cite{ramachandran08}.   While the spectral degree of freedom is by its nature continuous, we will show below that our implementation can lead directly to transverse mode - \emph{discrete} spectral hybrid entanglement.   Here, discrete spectral entanglement refers to the use of well-separated frequencies which are intrinsic to the SFWM process and which are not the result of projecting  a broader distribution onto certain frequencies.   Importantly, in our scheme the photon pairs are born in the nonlinear medium already with discrete hybrid entanglement in frequency and transverse mode, and do not require any additional preparation steps.  An important aspect of our source is that it permits control over the dimensionality of each of the two DOF, discrete frequency and transverse mode, as determined in part by the number of supported modes in the fiber.  An interesting further aspect that could be exploited in our source is that  each photon from a given pair in principle exhibits `single-photon entanglement' between frequency and transverse mode, which could be used for tests of realism and non-contextuality \cite{karimi10}.

\section*{Theory of intermodal SFWM}

From a standard perturbative analysis of SFWM in the presence of multiple transverse modes, the two-photon state is the coherent sum of contributions from all existing phasematched processes ($N$ below) and in general exhibits hybrid entanglement in frequency and transverse mode, as follows

\begin{equation}\label{E:state}
|\Psi\rangle=\sum\limits_{j=1}^{N} \eta_j\int\!\! d \omega_s \int\!\! d \omega_i  f_j(\omega_s,\omega_i) |\omega_s;\mu_j\rangle_s |\omega_i;\nu_j\rangle_i,
\end{equation}

\noindent where for each process $j$, $\eta_j$  and $f_j(\omega_s,\omega_i)$ represent the probability amplitude and  the joint spectral amplitude (JSA); $|f_j(\omega_s,\omega_i)|^2$ is referred to as the joint spectral intensity (JSI) .   Here, $|\omega;\mu_j\rangle_s$ ( $|\omega;\nu_j\rangle_i$)  represents a signal (idler) single-photon Fock state with frequency $\omega$ and spatial mode $\mu_j$ ($\nu_j$).  

Our goal is to generate photon pairs involving multiple SFWM processes, so that the JSA  for each process is factorable, i.e. so that it can be written as $f_j(\omega_s,\omega_i)=S_j(\omega_s)I_j(\omega_i)$.   In this case, the state which now exhibits discrete hybrid entanglement in frequency and transverse mode can be written as follows

\begin{equation}\label{E:stateshort}
|\Psi\rangle=\sum\limits_{j=1}^{N} \eta_j |\mu_j \rangle_{\omega_{s j}}|\nu_j \rangle_{\omega_{i j}},
\end{equation}

\noindent where $|\mu_j \rangle_{\omega_{s j}}$ and $|\nu_j \rangle_{\omega_{i j}}$ are signal and idler single-photon wavepackets, defined in terms of  the phasematched frequencies  $\omega_{s j}$ and $\omega_{i j}$, respectively, as

\begin{eqnarray}
|\mu_j \rangle_{\omega_{s j}} &\equiv \int d \omega S_j(\omega)|\omega;\mu_j \rangle,  \\
|\nu_j \rangle_{\omega_{i j}} &\equiv \int d \omega I_j(\omega)|\omega;\nu_j \rangle.
\end{eqnarray}

Note that if, in addition to each of the processes being factorable so that  Eq. \ref{E:stateshort} is valid, the sets of functions $\{S_j(\omega)\}$ and $\{I_j(\omega)\}$ are each orthogonal, then Eq. \ref{E:stateshort} actually represents a Schmidt decomposition.    If these sets of functions are non-overlapping, i.e if for any given $j$ the spectral region for which $S_j(\omega)$  is non-zero does not overlap similarly-defined regions for other $j's$ (and likewise for functions $\{I_j(\omega)\}$), it follows  that each set of functions must be orthogonal, and then each  $\{S_j(\omega),I_j(\omega)\}$ constitutes a Schmidt mode pair for the overall two-photon state, given as the coherent sum of the various phasematched processes.  Note that spectrally non-overlapping Schmidt mode pairs may alternatively be obtained through cavity-enhanced SFWM \cite{garay13, clemmen09, azzini12, Reimer14}  (or SPDC\cite{jeronimo10, Ou99}), for a sufficiently small pump bandwidth.

Consider a SFWM source based on a birefringent fiber~\cite{smith09,fang14, soller11} of length $L$ and on frequency-degenerate pumps, allowed to be non-degenerate in transverse mode, with a Gaussian spectral amplitude of bandwidth $\sigma$ centered at  $\omega_p$.  Assuming that each process $j$ is phasematched at signal (idler) frequencies $\omega_{sj}$ ($\omega_{ij}$), and defining detuning variables $\nu_\lambda \equiv \omega_\lambda-\omega_{\lambda j}$, with $\lambda=s,i$, we can write an expression for the JSA, correct to first order in $\nu_s$ and $\nu_i$, as $f_j(\nu_s,\nu_i)=\alpha(\nu_s,\nu_i)\phi_j(\nu_s,\nu_i)$ in terms of

\begin{align}
\alpha(\nu_s,\nu_i)&=\exp\left(-(\nu_s+\nu_i)^2/2\sigma^2\right) \label{alpha},\\
\phi_j(\nu_s,\nu_i)&=e^{-\frac{x^2}{D}}\!\! \left\{ \mbox{erf}(D/2-i x/D)+\mbox{erf}(i x/D)\right\} \label{phi},
\end{align}

\noindent where $\alpha(\nu_s,\nu_i)$ is referred to as the pump spectral amplitude and $\phi_j(\nu_s,\nu_i)$ is referred to as the phase matching function~\cite{garay07}.   Here,  $L(k_1+k_2-k_s-k_i) \approx T_{s} \nu_s+T_{i} \nu_i\equiv x$ is an adimensional phase mismatch and $\mbox{erf}(\cdot)$ is the error function, in terms of the following definitions 

\begin{align}
D&=\sigma L (k_1'-k_2')/\sqrt{2}, \\
T_{\mu}&=L[(k_1'+k_2')/2-k_\mu'] \ \ \  \mbox{with } \mu=s,i,
\end{align}

\noindent where a prime denotes a frequency derivative evaluated at $\omega_p$ for the pumps and at $\omega_{sj}$ ($\omega_{ij}$) for the signal (idler); $k_\lambda$ (with $\lambda=1,2,s,i$) is the wavenumber for each wave.  $D$ represents a non-degeneracy parameter, with $|\phi(x)|=\mbox{sinc}(x/2)$ at degeneracy ($D\rightarrow 0$). 

In order to reach a criterion for factorability, we approximate $|\phi_j(x)|$ as a Gaussian function $\mbox{exp}(-x^2/\Gamma^2)$; $\Gamma$ increases in value with $D$, with $\Gamma=4.55$ for $D \rightarrow 0$.    We impose the condition that  the term proportional to $\nu_s \nu_i$ in the argument of the combined exponential $|\alpha(\nu_s,\nu_i)\phi_j(\nu_s,\nu_i)|$ vanishes, thus obtaining the \emph{group velocity matching condition} (GVM)  $2 T_s T_i \sigma^2 /\Gamma^2=-1$~\cite{cohen09,halder09,soller11}.  This condition constrains: i) the fiber dispersion as $T_s T_i<0$, leading to  $k_s' \le k_p' \le k_i'$ or $k_i' \le k_p' \le k_s'$, with $k_p'\equiv(k_1'+k_2')/2$, and ii) the source parameters $L$ and $\sigma$.   The dispersion relation for transverse mode $l$, may be expressed as $k_l(\omega)=k_{mat}(\omega)+k_{wg,l}(\omega)$ in terms of the material and waveguide contributions,  $k_{mat}(\omega)$   and $k_{wg,l}(\omega)$; within the first-order dispersion approximation used  (see Eq. \ref{phi}), the state is fully determined by first derivatives of $k_l(\omega)$.  If the GVM condition is fulfilled for a specific combination of transverse modes and signal/idler frequencies, variations in the latter two will naturally contribute to a departure from this condition.  However,  if: i) the frequency spread between processes is sufficiently small and ii) the waveguide dispersion contribution is sufficiently weak, i.e.  $|k'_{wg,l}(\omega)/k'_{mat}(\omega) | \ll 1$, the GVM condition above can be fulfilled to good approximation by different processes, leading to simultaneous factorability across several processes.

\begin{figure}[t]
\centering
\includegraphics[width=16cm]{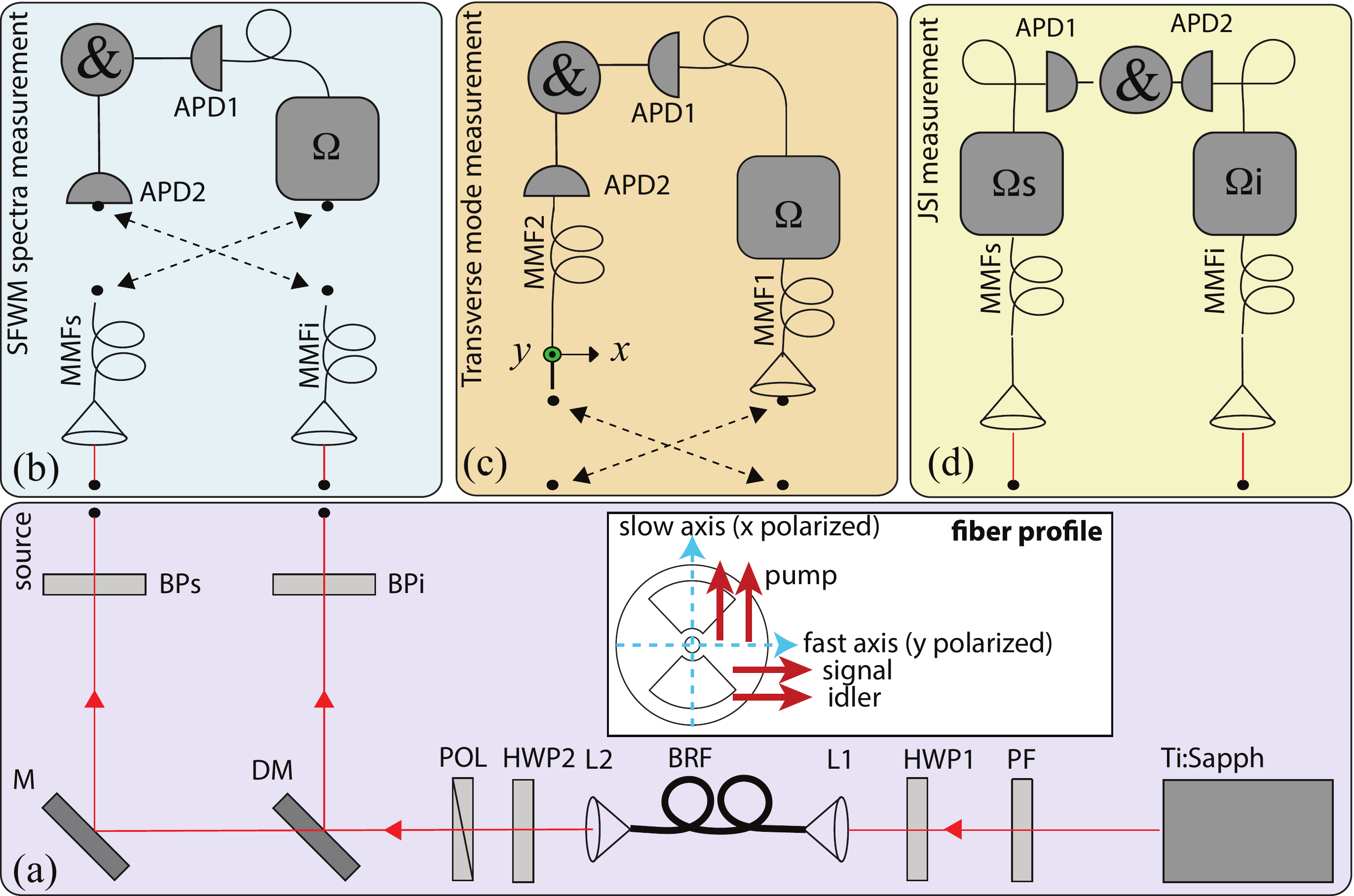}
\caption{\label{fig:setup} Experimental setup: (a) Implementation of cross-polarized SFWM in a bow-tie birefringent fiber (BRF); (b) Spectrally-resolved photon coincidence counting measurement, with spectral resolution for either the signal or idler mode ; (c) Spatially-resolved photon counting measurement; (d) Spectrally-resolved coincidence counting photon counting measurement, with spectral resolution for, both, the signal and idler modes.}
\end{figure}

\section*{Experimental implementation}

Our SFWM source, see Fig.~\ref{fig:setup}(a),  is similar to the one used in our earlier paper~\cite{cruz14} .  We employ as pump a picosecond mode-locked Ti:sapphire laser  (76MHz repetition rate and $0.52$nm bandwidth centered at  $690$nm, with spurious frequencies suppressed by a  band-pass filter, PF).  The pump beam (with $\sim50$mW power) is coupled into a $14.5$cm length of bow-tie birefringent fiber with an aspheric lens ($8$mm focal length; L1); a half-wave plate (HWP) sets the pump polarization for our cross-polarized SFWM process (see inset in Fig.~\ref{fig:setup}(a)) parallel to the fiber's slow axis. The photon pairs are out-coupled from the fiber using a lens, L2, identical to L1 and their polarization is set to horizontal using a second half wave plate (HWP2); a Glan-Thompson polarizer (POL) greatly reduces the remaining pump power.     The photon pairs are frequency non-degenerate, emitted in spectral bands which appear symmetrically around the pump;  they are split  using  a dichroic mirror (DM) followed by appropriate bandpass filters (BPs and BPi)  for further pump suppression.   

\begin{figure}[ht]
\centering
\includegraphics[width=12cm]{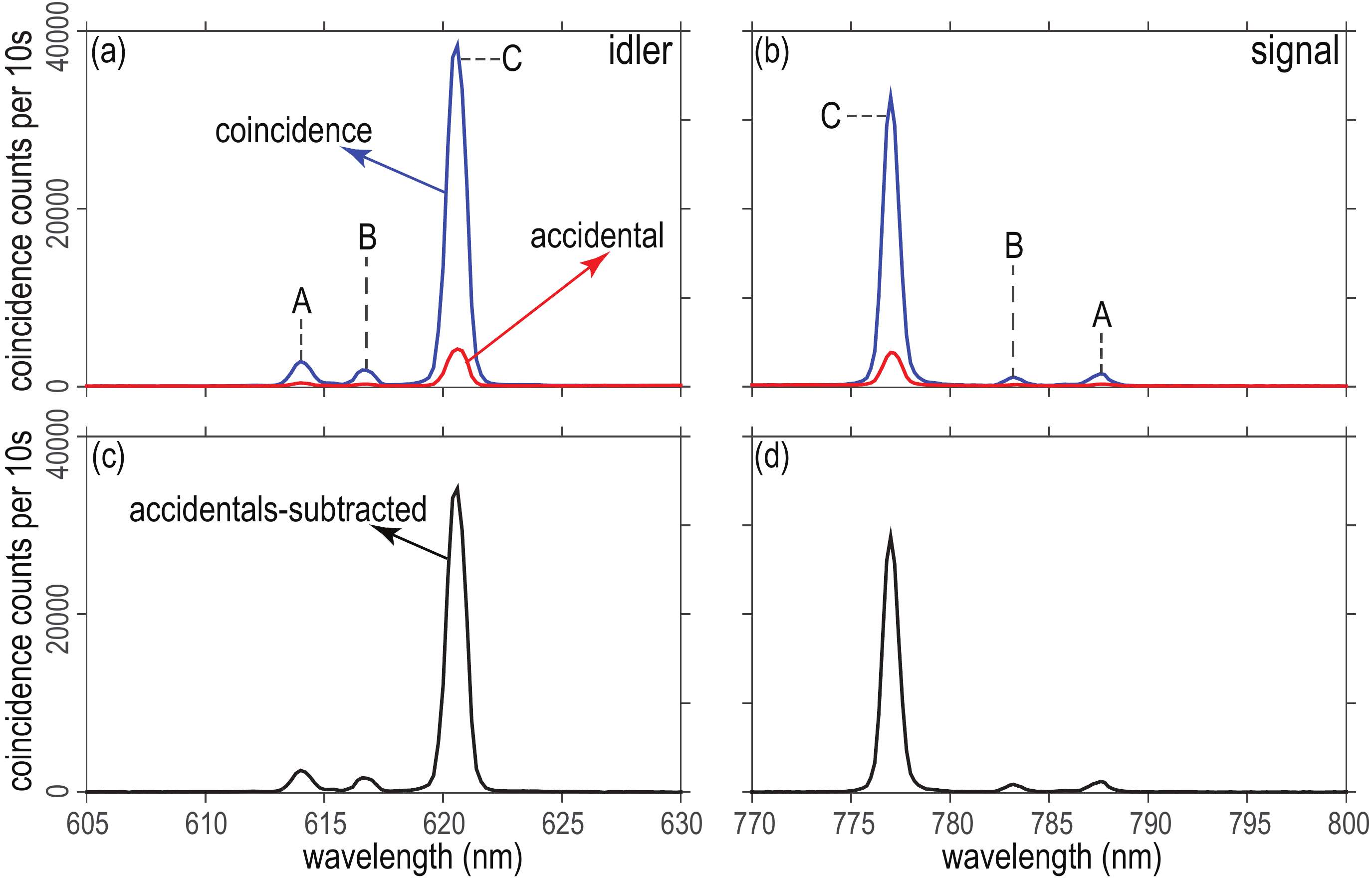}
\caption{\label{fig:peaks} Spectrally-resolved coincidence counts (blue curve) and accidental coincidence counts (red curve) for the idler mode, shown in panel (a), and for the signal mode, shown in panel (b). Coincidence counts with accidental counts subtracted for the idler mode, shown in panel (c), and for the signal mode, shown in panel (d).}
\end{figure}

\begin{figure}[ht]
\centering
\includegraphics[width=14cm]{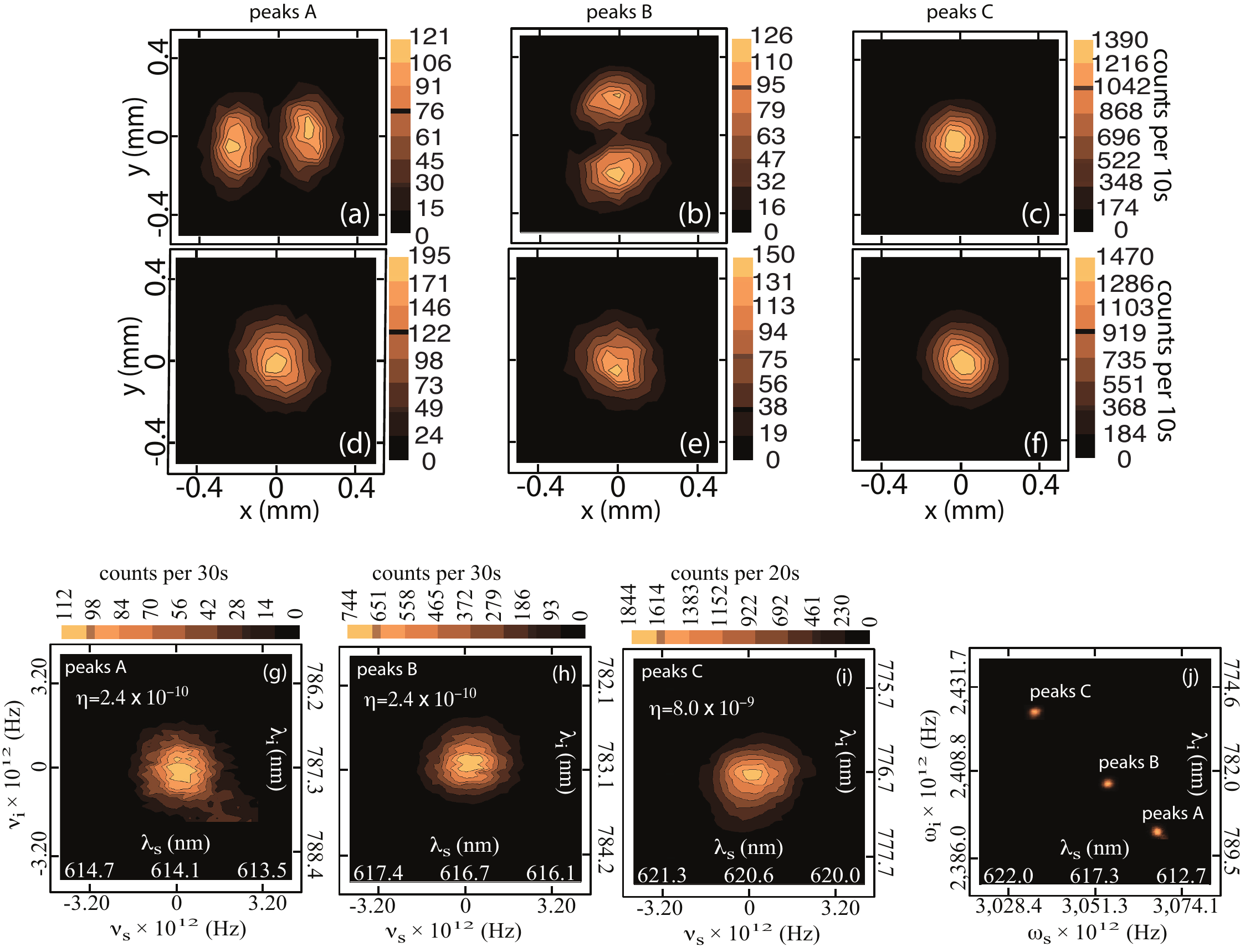}
\caption{\label{fig:exptalJSI}  Spatially-resolved transverse intensity corresponding to each of the six peaks in Fig.\ref{fig:peaks} for the idler [signal] shown in (a)-(c) [(d)-(f)]. Experimentally-measured JSI for peaks $A$, $B$, and $C$ shown in panels (g)-(i), indicating experimental estimated process efficiencies $\eta$ (removing the effect of losses); while the spectral windows have essentially identical dimensions to those in Fig.~\ref{fig:jointspectra} (d),(g),(j), the slight shift is probably due to calibration error in the monochromators. All three JSI's shown together in panel (j).}
\end{figure}

We employ three different detection schemes, relying on a combination of spectrally- and spatially-resolved photon counting.  First, we characterize the spectral structure of the photon pairs; see Fig.~\ref{fig:setup}(b). We couple 
the idler  ($\lambda<\lambda_p$) and  signal  ($\lambda>\lambda_p$) photons into  separate multimode fibers, MMFi and MMFs.  A grating monochromator ($\Omega$) takes MMFs as input and its output is directed to  a silicon avalanche photodiode (APD1);  MMFi leads directly to a second detector (APD2). We scan the spectral transmission window of $\Omega$, while monitoring  the coincidence rate at APD1 and APD2.   We are thus able to measure the signal-photon spectrum in coincidence with the corresponding (spectrally-unresolved) idler photon; subsequently, we reverse the roles of the two photons so as to measure the idler-photon spectrum.  In Fig.~\ref{fig:peaks} (a) and (b),  the blue curves show our measurement of the spectrally-resolved coincidence counts for the idler mode in panel a, and for the signal mode in panel b.  The red curves show the corresponding accidental coincidence counts obtained by repeating the previous measurement, except with an added relative delay, equal to the time interval between pump pulses, between the signal and idler channels. In Fig.~\ref{fig:peaks} (c) and (d) we show the spectrally resolved coincidence counts with accidentals subtracted, for the idler and signal modes, respectively.   It is apparent that the detection events appear in three pairs of energy-conserving peaks, labeled as $A$, $B$ and $C$.


In a second measurement,  see Fig.~\ref{fig:setup}(c),  we verify that each measured peak corresponds to a well-defined transverse mode.   We couple the idler  photon into a multimode fiber (MMF1), leading to the entrance port of a monochromator ($\Omega$)  and successively center its spectral transmission window at each of the three peaks in Fig.~\ref{fig:peaks}(c);  the output of $\Omega$ is directed to detector APD1. In addition, with the help of lens L2, the output plane of BRF is imaged to a plane along the signal arm, on which we place the tip of a multimode fiber (MMF2) leading to detector APD2.  The fiber tip of MMF2 can be displaced on a transverse plane with computer-controlled motors, so that the coincidence rate between APD1 and APD2 as a function of the fiber tip's position reveals the transverse intensity of a signal photon corresponding  to a specific frequency as set by the spectrally-resolved detection of the idler photon.   Subsequently, the roles of the photons are reversed, so as to measure the transverse intensity of the idler photon.

Figs.~\ref{fig:exptalJSI}(a) through (c) show measurements of the idler single-photon transverse-intensity distribution, as triggered by the detection of a signal photon at each of the three available  peaks.   Figs.~\ref{fig:exptalJSI}(d) through (f) show corresponding measurements for the signal photon.    Note that the modes measured and shown in Fig.~\ref{fig:exptalJSI}(a-f) match those for processes $a$, $b$, and $c$ (see discussion below for the definition of these three processes). In accordance with Eq. \ref{E:state},  this type of source produces photon pairs which in general exhibit hybrid entanglement in discrete frequency and transverse mode.    The resulting type of entanglement is determined by the phasematching characteristics which in turn are governed by the specific fiber dispersion experienced by the four participating waves.   For our specific experimental configuration while each idler peak corresponds to a different transverse mode, all signal peaks correspond to the fundamental mode implying that there is no entanglement in transverse mode.  The reason for this is that only the fundamental mode is supported for the signal wave (which propagates with longer wavelengths than the pump); this clearly implies that all permitted processes must involve the fundamental mode for the signal photon, thus precluding the presence of spatial entanglement.   Note that, in general, the state becomes richer with more phasematched processes as the core radius is increased and/or the pump wavelength is reduced, so that more modes are supported for some or all of the four waves involved.   As we will show below, reducing the pump wavelength to $620$nm while retaining the same fiber, leads to ten phasemached processes exhibiting a rather rich hybrid entanglement structure.   This interesting configuration, however, is currently out of reach from our experimental capabilities since our laser cannot be tuned to $620$nm.

We have carried out a third measurement, the JSI, for each of the three pairs of peaks; see Fig.~\ref{fig:setup}(d).  We couple the idler and signal photons into multimode fibers (MMFi and MMFs), which lead to the entrance ports of two separate monochromators ($\Omega_i$ and $\Omega_s$);   the outputs of $\Omega_s$ and $\Omega_i$ are directed towards two detectors APD1 and APD2.   The coincidence count rate in APD1 and APD2 is recorded as a function of the central transmission frequencies of  $\Omega_s$ and $\Omega_i$.  This measurement is carried out around the phasematched frequencies $\omega_{sj}$ and $\omega_{ij}$, for each of the three pairs of energy-conserving peaks.  The results are presented in Fig.~\ref{fig:exptalJSI}.   While panels (g) through (i) show the JSI measurement for each of pairs A through C, panel (j) shows the three measurements in a combined plot.   Note that there is good agreement between the measured JSI's (Fig.~\ref{fig:exptalJSI}) and the calculated ones (Fig.~\ref{fig:jointspectra}).  

\section*{Discussion}

In order to analyze our experimental results, we consider a fiber that supports the linearly polarized modes $\mbox{LP}_{01}$ and $\mbox{LP}_{11}$;  fiber birefringence then implies the existence of two non-degenerate $\mbox{LP}_{01}$  modes: $\mbox{LP}_{01x}$ and $\mbox{LP}_{01y}$, polarized along the $x$ and $y$ directions, and four non-degenerate $\mbox{LP}_{11}$  modes: $\mbox{LP}_{11ex}$, $\mbox{LP}_{11ey}$, $\mbox{LP}_{11ox}$, and $\mbox{LP}_{11oy}$, with even($e$)  and odd($o$) parity~\cite{modeunfolding}.   We have used a simple dispersion model where, for polarization $i=x,y$ and parity $j=e,o$, the index of refraction is given as $n_{ij}(\omega)=n(\omega)+ \delta_{i x} \Delta+ \delta_{j o} \Delta_p$, where $\delta_{ij}$ is a Kronecker delta; here, $n(\omega)$  is obtained for a step-index fiber characterized by radius $r$ and numerical aperture $NA$, while $\Delta$ is the birefringence and $\Delta_p$ is the ``parity  birefringence''~\cite{garay15} (with $\Delta_p=n_{xo}-n_{xe}=n_{yo}-n_{ye}$).

\begin{figure}[t]
\centering
\includegraphics[width=16cm]{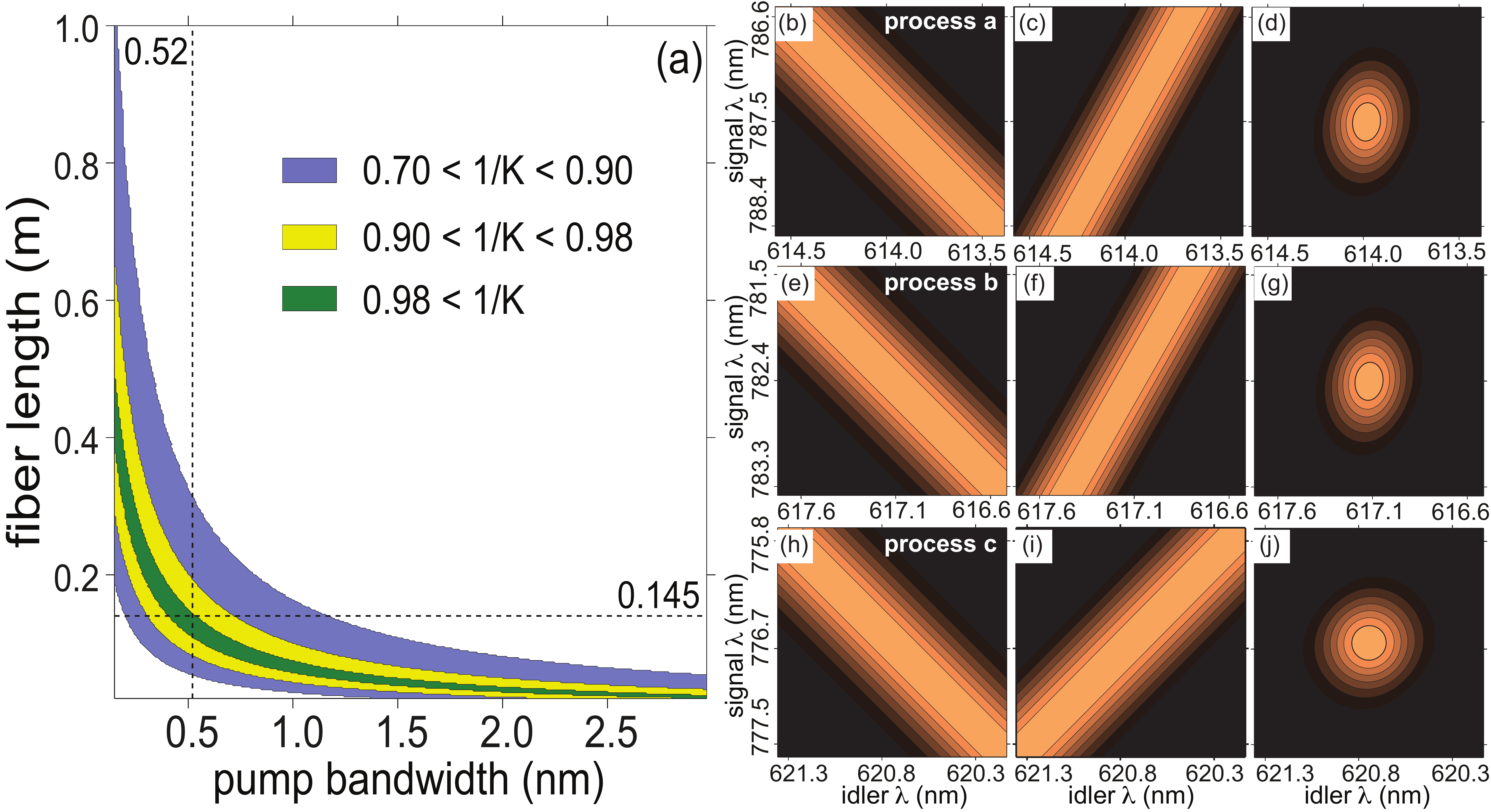}
\caption{\label{fig:jointspectra} (a) Regions of simultaneous near-factorability for all three processes, (b)-(j): Plots of $|\alpha(\nu_s,\nu_i)|^2$,  $|\phi_j(\nu_s,\nu_i)|^2$, and $|\alpha(\nu_s,\nu_i)\phi_j(\nu_s,\nu_i)|^2$ for processes $a$ [(b)-(d)], $b$ [(e)-(g)], and $c$ [(h)-(j)].}
\end{figure}

We concentrate on cross-polarized SFWM processes, for which the pump is $x$-polarized, and the SFWM photons are $y$-polarized. The fiber birefringence then implies a shift $\delta=2 \omega_p \Delta/c$ in the phasematching condition $\Delta k^{||}+\delta=0$, where $\Delta k^{||}$ is the phasemismatch for an equivalent co-polarized process.  The essential advantage of cross-polarized SFWM is that this shift can be large enough for $\omega_s$ and  $\omega_i$ to be sufficiently removed from $\omega_p$  to avoid Raman contamination, while the GVM conditions $k_s' \le k_p' \le k_i'$ or $k_i' \le k_p' \le k_s'$  tend to be fulfilled naturally, provided that  $\omega_p$, $\omega_s$, and $\omega_i$ are not in the vicinity of a zero dispersion frequency (so that $k'$ has the same sign for all three waves).

Let us note that with $M$ supported modes, in principle $M^4$ processes are possible, which in our case ($M=6$) translates into $1296$ processes.   As we discussed in Ref.\cite{garay15}, conservation of parity and orbital angular momentum constraints may be used to determine the subset of viable processes.  In our specific case, considering only cross-polarized processes of the type $xx-yy$,  $15$ processes are viable; in order to determine which of these $15$ processes actually take place we implement the following strategy.  In the experiment we measure the SFWM spectrum, which comprises three pairs of energy-conserving peaks, labelled $A$, $B$ and $C$, and for each peak we measure the transverse intensity which corresponds to each peak (see Fig.~\ref{fig:exptalJSI}, above), so that in effect we directly observe the transverse mode corresponding to each of the six spectral peaks.  We input the center frequency and transverse mode for all six peaks into a genetic algorithm (GA) in order to search for combinations of parameters $\{r,NA,\Delta,\Delta_p\}$ that best explain our results;   our GA is based on a  fitness function defined as the sum of the absolute value of the phasemismatch functions for the three pairs of peaks, evaluated at its center frequencies~\cite{garay15}.

For our specific ``bow-tie'' fiber (HB800C from Fibercore Ltd) we obtain values $r=1.74\mu$m, $NA=0.17$, $\Delta=2.37\times10^{-4}$, and $\Delta_p=4.41\times10^{-4}$ which match well those provided by the manufacturer (except $\Delta_p$ which is not specified).  From the minimization of the fitness function in our GA, we also determine that such a fiber supports three processes when pumped at $690$nm, $a$, $b$, and $c$,  each one involving a specific combination of transverse modes for the pump 1, pump 2, signal and idler waves:   $\mbox{LP}_{01x},\mbox{LP}_{11ex}, \mbox{LP}_{01y}, \mbox{LP}_{11ey}$ (process $a$), $\mbox{LP}_{01x}, \mbox{LP}_{11ox}, \mbox{LP}_{01y}, \mbox{LP}_{11oy}$ (process $b$), and $\mbox{LP}_{01x},\mbox{LP}_{01x},\mbox{ LP}_{01y}, \mbox{LP}_{01y}$ (process $c$), with the property that peaks $A$ result from process $a$, peaks $B$ result from process $b$, and peaks $C$ result from process $c$.

As discussed in Ref. \cite{garay15}, the efficiency of each process is determined by an overlap integral between the transverse modes for each of the four participating waves.    Because for  process $c$ all four waves propagate in the fundamental LP$_{01}$ mode, the degree of overlap is significantly higher than for processes $a$ and $b$, which involve both the LP$_{01}$ and  LP$_{11}$ modes.   This explains the fact that process $c$ leads to approximately one order of magnitude higher flux than processes $a$ and $b$, despite all three processes being perfectly phasematched at the corresponding center frequencies.    It is important to point out that besides the overlap integral, the resulting flux is also dependent on the fraction of the pump power that couples into each of the (pump) modes supported.   Indeed, the probability amplitude for process $j$, $\eta_j$ (see Eq. \ref{E:state}), is proportional to the product of the electric field amplitudes for the two participating pumps in the SFWM process in question.  This means that the relative flux emitted by the various processes may be controlled in principle through the relative intensities coupled into each of the pump transverse modes.  For the particular case of our source, pump mode LP$_{01x}$ participates in all three processes, but mode LP$_{11ex}$ participates only in process $a$, and mode  LP$_{11ox}$ participates only in process $b$.   Thus, increasing the fraction of pump power coulpled into mode  LP$_{11ex}$ will boost the flux of process $a$  at the expense of the other two, and likewise increasing the fraction of pump power coupled into mode  LP$_{11ox}$ will boost the flux of process $b$. Additionally, control over the relative phases of the pump modes, prior to fiber coupling, would translate into control over the relative \emph{phases} of  the various SFWM processes.  Such control over the amplitude and phase of the various pump modes may in principle be accomplished by preparing in free space the desired specific combination of modes to be coupled into the fiber, for example with the help of a spatial light modulator \cite{Bouchal05}.  Note that in order to maintain the phase relationships in the two-photon state, it is interesting to consider the use of fibers specifically designed for the suppression of coupling between transverse modes \cite{ramachandran08}.   This would permit the faithful transmission of our photon pairs with hybrid entanglement over long distances. 

The degree of (spectral) factorability for each process can be quantified through the reciprocal Schmidt number $K^{-1}$ of the two-photon state, appropriately filtered so that the other two processes are suppressed.  We  calculate the expected Schmidt number by carrying out a numerical singular value decomposition of the theoretical joint amplitude, using the experimental parameters in the previous paragraph. Note that $K^{-1}$  is also the purity of a single photon (say the signal), heralded by the detection of its conjugate (say the idler).   While $K^{-1}=1$ indicates perfect factorability, $K^{-1} \rightarrow 0$ indicates maximal entanglement.    In  Fig.~\ref{fig:jointspectra}(a) we show the regions in  $\{\Delta\lambda_p, L\}$ space ($\Delta\lambda_p\propto\sigma$) for which \emph{all three processes} obey the GVM condition and are therefore nearly-factorable with  $K^{-1}>0.7$ (shown in blue),  $K^{-1}>0.9$ (yellow), and $K^{-1}>0.98$ (green).  Note that non-ideal values $K^{-1}<1$ for one or more processes would imply that the individual joint amplitude functions would no longer constitute independently-addressable (via spectral filtering) Schmidt mode pairs, as is desired for controllable scalability to higher dimensions.  In Fig.~\ref{fig:jointspectra}(a) we have indicated the specific values of the fiber length and the pump bandwidth used in the experiment, $L=14.5$cm and $\Delta\lambda_p=0.52$nm (shown with dotted lines), which as can been seen, lead to all three processes being nearly-factorable.  In each row of Fig.~\ref{fig:jointspectra} (b)-(j)  we show, plotted vs $\omega_s$ and $\omega_i$,  the functions $|\alpha(\nu_s,\nu_i)|^2$, $|\phi_j(\nu_s,\nu_i)|^2$, and $|\alpha(\nu_s,\nu_i) \phi_j(\nu_s,\nu_i)|^2$, for processes $a$, $b$, and $c$; the negative slope for $|\alpha(\nu_s,\nu_i)|^2$ \emph{and} positive slope for $|\phi_j(\nu_s,\nu_i)|^2$, 
translates, for the specific widths of these functions (controlled by $\sigma$ and $L$ respectively), into a nearly factorable JSI for all three processes.  

\begin{figure}[t]
\centering
\includegraphics[width=16cm]{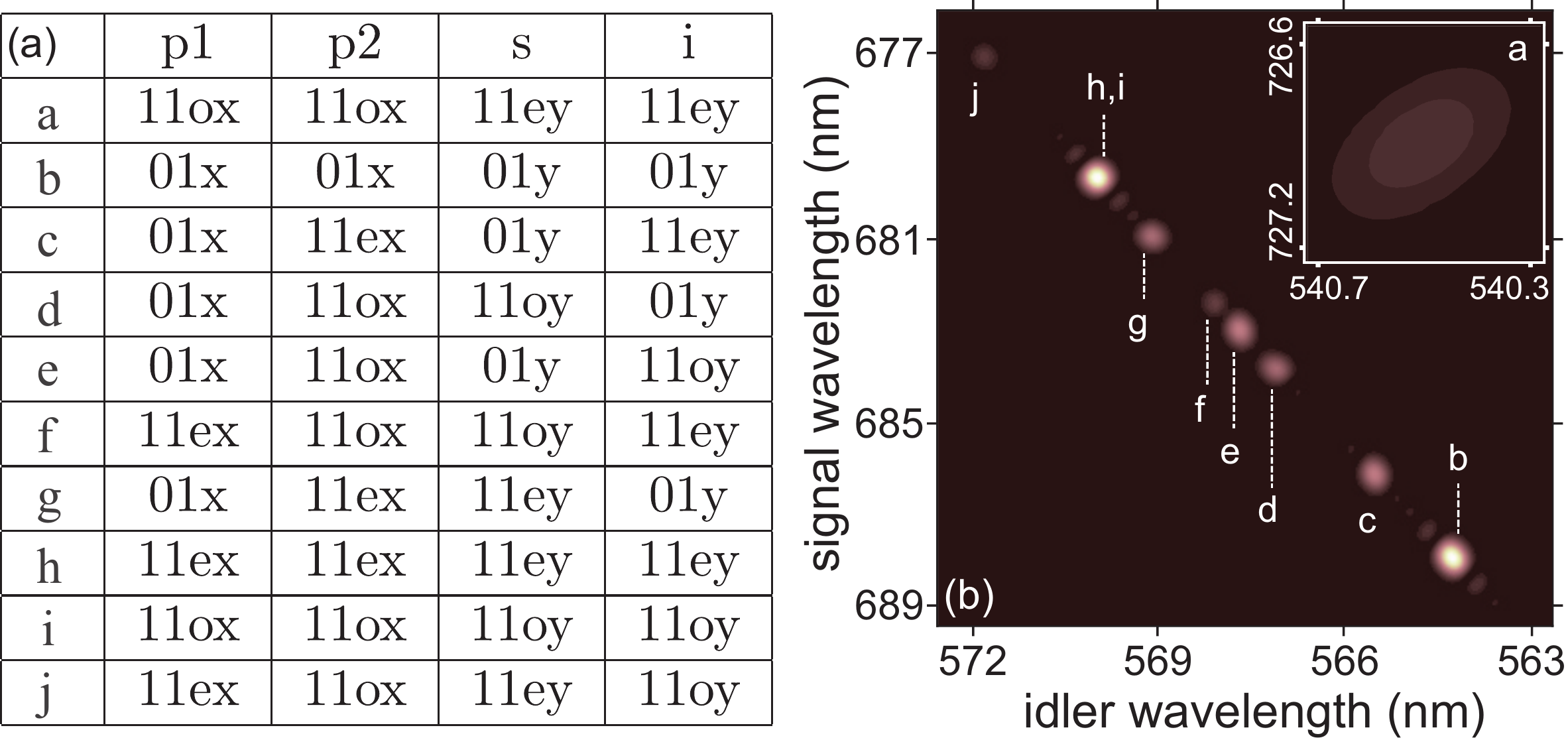}
\caption{\label{fig:scalability} Table: Participating transverse LP modes  in each of 10  processes ($a$-$j$). Figure: JSI functions for all ten processes; note that process $a$ (shown in inset) is spectrally distant from processes b-j  and does not fulfil the GVM condition.}
\end{figure}

Note that the number of SFWM processes (and therefore the entanglement dimensionality) can be scaled up as a result of additional mode combinations made available if the number of  supported modes  in the fiber is increased for some or all of the the four waves involved.   While in the source configuration used in the experiment, the fiber only supports the fundamental mode for the range of wavelengths corresponding to the signal photon, the situation can change if we reduce the pump wavelength or increase the core radius.   The effect of using a lower pump wavelength, so that the fiber supports the $LP_{11ey}$ and $LP_{11oy}$ modes for the signal photon,  is illustrated in Fig.~\ref{fig:scalability} in which we show the  simulated joint spectra corresponding to ten distinct processes which now occur.  For this simulation we have  assumed the same fiber as used in our experiments, with a pump wavelength of $620$nm, under the assumption that the pump power is coupled with equal efficiencies to the modes $LP_{01x}$,  $LP_{11ex}$, and $LP_{11ox}$.   In the table, we indicate the transverse modes which participate in each of the ten resulting processes.  Note that we select the fiber length and pump bandwidth values,   $L=12$cm and $\Delta \lambda=0.35$nm, so as to attain near simultaneous factorability across these processes (process $a$, which is considerably distant from the other nine processes in the joint frequency space, ends up being less factorable). Unfortunately, this source configuration is not feasible in our setup, since $620$nm lies outside of the tuning range of our Ti:sapphire laser.   Nevertheless, this figure serves as illustration of  how the additional processes made possible when the fiber supports a larger number of transverse modes for some or all of the waves involved (in this case for the signal photon),  leads to a rather rich hybrid entanglement structure.   The dimensionality associated with the two degrees of freedom, frequency and transverse mode, clearly is scaled up as the state becomes more complex (in this case as a consequence of a reduction in the pump wavelength).   As can be seen from Fig.~\ref{fig:scalability} processes $b$, $h$, and $i$ which corresponds to all four waves propagating in modes with an essentially identical spatial profile, are the most efficient because of their superior mode overlap.    Note that processes $h$ and $i$ overlap spectrally; by post-selecting these two processes with appropriate spectral filters, it would be possible to generate a state entangled in transverse mode, of the sort $r | 11ey\rangle_{\lambda s} | 11ey\rangle_{\lambda i}+s| 11oy\rangle_{\lambda s}| 11oy\rangle_{\lambda i}$ with $\lambda_s \approx 679.7$nm and $\lambda_i \approx 570.0$nm.  This can become a Bell state for balanced coefficients $ |r |= |s |=1/\sqrt{2}$, opening an interesting route to the generation of spatially-entangled photon pairs.

\section*{Conclusions}

In conclusion, we have designed and implemented a photon-pair source, based on the SFWM process in a few-mode birefringent fiber.  The two-photon state produced by this source includes contributions from three separate SFWM processes, each corresponding to a phasematched combination of transverse modes and frequencies of the participating waves (pumps 1 and 2, signal, and idler).   The state produced by a source of this kind in general exhibits hybrid entanglement in discrete frequency and transverse mode, and the dimensionalities associated with each degree of freedom may be controlled through the pump wavelength and/or fiber core radius.   Furthermore, we have shown that the joint spectral intensities for the three separate processes  may be made nearly-factorable through an appropriate choice of the fiber length and the pump bandwidth.   In order to characterize our source we have measured for each process:  i) the SFWM spectrum (in coincidence counts), leading to three pairs of energy conserving peaks, ii)  the transverse spatial mode on the output plane of the SFWM fiber, for each of the two corresponding  peaks, and iii) the joint spectral intensity.   We believe that our work will open up a new path towards controlled scalability  to higher-dimensional hybrid entanglement in photon pairs.

\section*{Acknowledgements }

This work was supported by CONACYT (grants 230072, 222928, 253366, and 221052) and by PAPIIT(UNAM) grant IN1050915.

%

%

\end{document}